\documentclass[%
 reprint,
 amsmath,amssymb,
 aps,
floatfix,
float,
]{revtex4-1}

\newcommand{\RNum}[1]{\uppercase\expandafter{\romannumeral #1\relax}}

\usepackage{graphicx}
\usepackage{caption}
\usepackage{float}
\usepackage{xcolor}
\usepackage[caption=false]{subfig}
\usepackage{dcolumn}
\usepackage{bm}
\usepackage{hyperref}

\newcounter{eqn}

\DeclareMathOperator*{\argmin}{argmin}

\makeatletter
\newcommand{\putindeepbox}[2][0.7\baselineskip]{{%
    \setbox0=\hbox{#2}%
    \setbox0=\vbox{\noindent\hsize=\wd0\unhbox0}
    \@tempdima=\dp0
    \advance\@tempdima by \ht0
    \advance\@tempdima by -#1\relax
    \dp0=\@tempdima
    \ht0=#1\relax
    \box0
}}
\makeatother

\begin{document}

\title{Pattern phase diagram of spiking neurons on spatial networks}

\author{Dionysios Georgiadis}
\affiliation{%
 	Future Resilient Systems, Singapore}
\affiliation{%
	Department of Management, Technology and Economics, ETH Zurich, Switzerland\\}
\author{Didier Sornette}
\affiliation{%
	Department of Management, Technology and Economics, ETH Zurich, Switzerland\\}

\date{\today}

\begin{abstract}
We study an abstracted model of neuronal activity via numerical simulation, and report spatiotemporal pattern formation and critical like dynamics.
A population of pulse coupled, discretised, relaxation oscillators is simulated over networks with varying edge density and spatial embeddedness.
For intermediate edge density and sufficiently strong spatial embeddedness, we observe a novel spatiotemporal pattern in the field of oscillator phases, visually resembling the surface of a frothing liquid. 
Increasing the edge density results in critical dynamics, with the distribution of neuronal avalanche sizes following a power law with exponent one. 
Further increase of the edge density results in metastable behaviour between pattern formation and synchronisation, before transitioning the system entirely into synchrony.
\end{abstract}

\maketitle


\section{Introduction}

Pulse coupled oscillator models (PCO) are defined as populations of relaxation oscillators, interacting in a pulse-like manner over some topology.
Such models have provided insight into the phenomenon of spontaneous synchronisation across fields, be it swarms of blinking fireflies,  pulsing pacemaker heart cells\cite{mirollo1990synchronization}, distributed computing systems\cite{proskurnikov2017synchronization,konishi2008synchronization}, or traders in financial markets \cite{wray2014cascades,wray2016financial}. 
PCOs have been prominent in neurology, since they can demonstrably capture multiple features of the rich dynamic behaviour of biological neuronal systems, such as: metastability (with the same configuration alternating between synchronous and asynchronous behaviour)\cite{deville2008synchrony}, spatiotemporal pattern formation (in the form of nonlinear waves)\cite{huang2004spiral,wilson1972excitatory,deville2005two,ermentrout2001traveling,Guardiola99}, and critical dynamics\cite{deville2008synchrony,bottani1995pulse,friedman2012universal}. 

Models used in this literature range from intricate (as for example the Hodgkin-Huxley model), to relatively simple (such as Integrate \& Fire oscillators). 
Deville and Peskin  \cite{deville2008synchrony} studied a population of Discretised Integrate \& Fire (DIF) oscillators, arguably the most abstracted model of neuronal activity to date.
In spite of its simplicity, the DIF model with all-to-all stochastic interactions can successfully capture metastability and synchrony \cite{deville2008synchrony}, as observed in cortical networks. 
However, the extend to which the DIF can replicate the phenomenology of its more intricate counterparts is not well understood, which is the motivation for the present paper.

Numerical studies have considered the DIF over quenched nontrivial topologies, focusing either on synchrony in complex networks\cite{deville2012synchrony} or on replicating temporal features of biological neuronal networks \cite{friedman2012universal}.
Notably, in \cite{friedman2012universal}, populations of cascading spiking neurons in lab-grown cortical slices were found to follow a power law distribution as well as a specific temporal profile. 
Both these features were captured by a single simulation of a DIF model over a regular lattice, revealing that critical-like behaviour is possible in DIF models.
This observation has not been studied further - in spite of the recent, sharp interest in models of criticality in neuronal systems\cite{beggs2003neuronal,kinouchi2006optimal,beggs2008criticality}.
Furthermore, pattern formation in DIF models has so far not been studied at all.

To address these two issues, we consider the DIF over topologies ranging between two purposefully picked extremes: 
a random graph (that has been studied the most in the literature so far) and a spatial graph (which does not disregard the spatial nature of biological systems).
Our results showcase that spatial embeddedness endows the DIF model with pattern forming behaviour and critical like dynamics, and therefore the unrealistic all-to-all annealing topology drastically restricts the phenomenology of the DIF.
Specifically, for sufficiently low spatial embeddedness, the field defined by the oscillator phases forms a novel spatiotemporal pattern - visually reminiscent of the surface of a frothing liquid.
Further increasing the number of connections between oscillators, we report metastability between pattern formation and synchrony (a behaviour of biological neuronal networks\cite{huang2004spiral}), while the transition itself is characterised by critical-like dynamics.
For even higher edge density, near periodic synchronous firing of all neurons ensues.

\section{The model}

Consider $N$ identical oscillators over a biderctional unweighted graph $\mathcal{G}$. Each node is associated with a binary \textit{state} $a_i$ ($1$ fired, $0$ not fired) and  a discrete, non-negative \textit{phase} $\phi_i$. We initialise all oscillator phases uniformly at random, and all states at zero. We then apply \textit{stochastic drive}, by randomly picking a small population of $d$ oscillators and increasing their phase by one unit. Variable $d$ is dubbed \textit{drive rate}.

\paragraph{Dynamics:}
 Once the phase of an oscillator reaches the \textit{threshold} value $\Theta$ the oscillator is said to \textit{fire}: its state is set to one, and the phases of all its neighbours are increased by one.
 This pulse-like interaction results in avalanching events, which we will refer to as \textit{cascades}. 
 The cascade continues as long as new oscillators fire under the influence of their neighbours. Crucially, an oscillator is \textit{only allowed to fire once} per cascade - a property dubbed refractoriness. 
 This is realised in the model through the state $a_i$: an oscillator may only fire if its state starts from zero. 
 Thus, once the oscillator fires, and its state is set to one, it is unable to fire again. 
 The state remains one until the cascade ceases. 
 Then, all fired oscillators are reset: $(a_i,\phi_i)\leftarrow (0,0), ~ \forall ~\{i:a_i=1\}$. After resetting all fired oscillators, we resume the stochastic drive, until a new cascade starts. The number of oscillators that participated in the cascade is \textit{the cascade size}.

\paragraph{Topology:}
The graph of interactions $\mathcal{G}$ is a geometric random graph\cite{penrose2003random}, furnished with random long-range connections. The random geometric graph is arguably the most parsimonious model of spatially embedded networks, constituting a canonical choice for the current study. Given the number of nodes $N$, the average degree $E$ and the fraction of long-to-short range connections $R$, the topology is assembled in the following way:
\begin{enumerate}
\item{`Sprinkle' $N$ points uniformly at random over a one-by-one square space.}
\item{Find the \textit{unconnected} pair of points with the minimum Euclidean distance and connect it. Repeat, until $E(1-R)$ edges have been drawn. We will refer to these edges as \textit{short range connections}.}
\item{Pick $ER$ random pairs of nodes and connect them, forming the \textit{long range connections}.}
\end{enumerate}
For extremely low values of $R$, we observe a heavily `meshed' structure, with a large topological diameter. 
For $R$ close to one, we retrieve a random graph with zero spatial embeddedness. In the current work, periodic boundary conditions were used in the making of $\mathcal{G}$.

\paragraph{Additional nomenclature:} 
We measure time in discrete `long' time, where an additional cascade corresponds to an additional `long' time unit. This is to be contrasted with the `short' time, which is associated with the dynamics within a cascade.
A total of $n_s$ cascades are simulated, and the fractional size the $t$th cascade is denoted by $c_t$ (with $c_t=0$ if no cascading occured). 
Additionally, the phases $\phi_i$ form a field $\Phi_t = ( \phi_0, \dots ,\phi_N )$.
Both $c_t$ and $\Phi_t$ vary in time, forming timeseries: $C=( c_0 ,\dots c_{n_s}, )$ and $\Phi=( \Phi_0 ,\dots ,\Phi_{n_s} )$. 
Both $C$ and $\Phi$ are dependent on the parameters $E,R,\Theta,N$. However, for the sake of notational simplicity, these dependences will remain implicit unless necessary.  

\paragraph{Simulation parameters:} Throughout the study, we fix $\Theta=5,~d= N/{10^3}$. Unless otherwise stated, we always simulate $5 \cdot 10^4$ cascades, discarding the first $10^4$ events - to ensure that the dynamics have reached stationarity. This results in $n_s=4 \cdot 10^4$. $R,E,N$ will all be specified for each experiment.

\section{Identifying regimes }

For sufficiently high mean degree $E$, we observe cascades of scale $\mathcal{O}(N)$ over regular time intervals, a property to which we will refer to as \textit{near periodic synchrony}.
We quantify this behaviour with the help of the metric $h \in [0,1]$, which is formally defined as the normalised Herfindahl index of the temporal power spectrum of $C$, the time series of cascades on the `long' time process.
Values of $h$ near one correspond to asynchrony, while low values of $h$ imply synchrony. For the sake of succinctness, details on this method can be found in Appendix A.

\begin{figure*}[!htp]
\center
\includegraphics[width=1.94\columnwidth]{"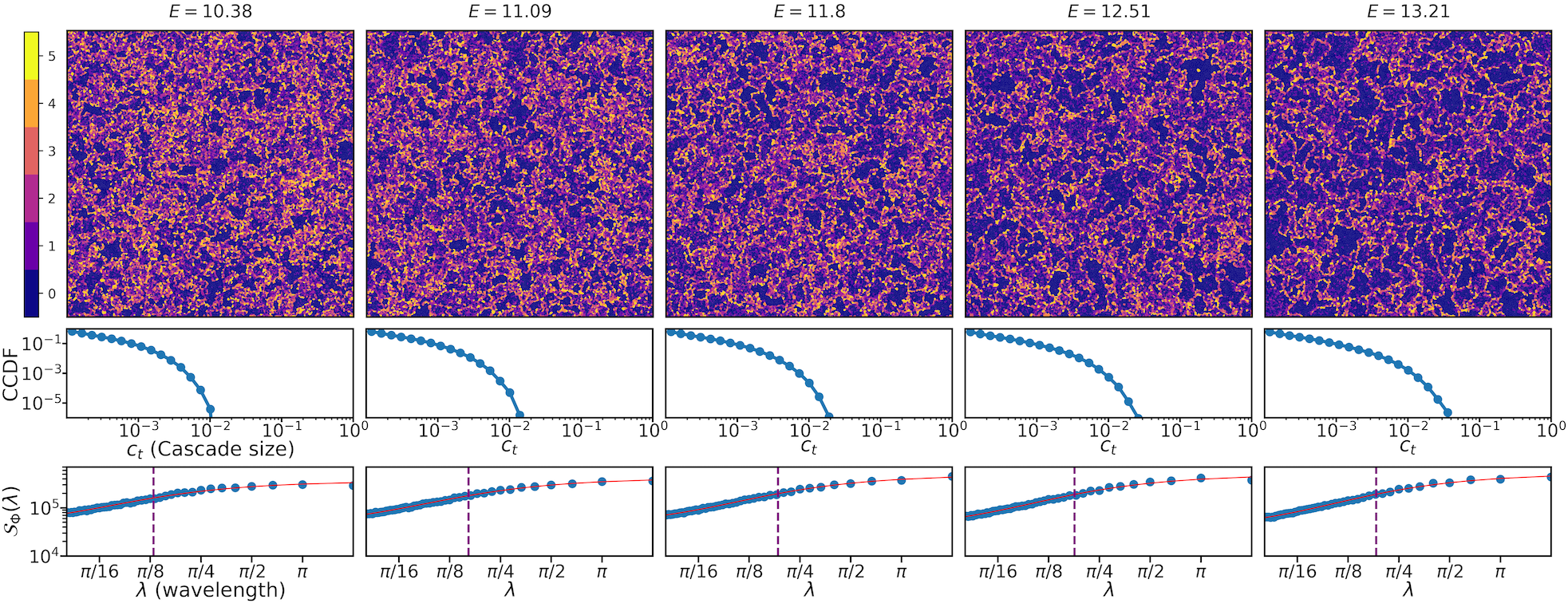}\\
\includegraphics[width=1.94\columnwidth]{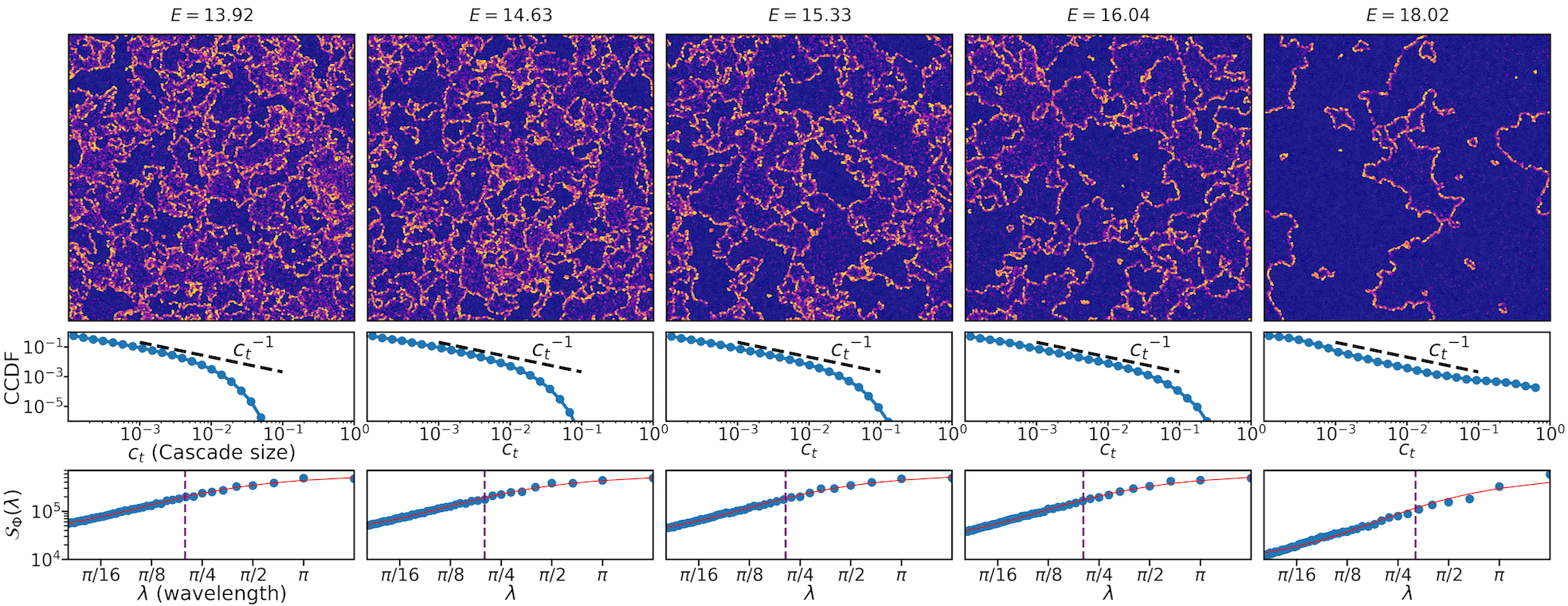}
\caption{Rows 1 and 4: snapshots of the oscillator phase fields $\Phi_t$. The scale of the patterns $\chi$ increases along with the mean degree $E$. 
Rows 2 and 5: CCDF of cascade sizes for the simulations of the row above. 
As $\chi$ increases, the CCDF extends towards the left - at first forming a power law and eventually a supercritical distribution (bottom rightmost panel).
Rows 3 and 6: blue dots correspond to the temporal average of the spatial spectral power $\mathcal{S}_{\Phi}(\lambda)$ defined in \eqref{mathcals}. 
The solid red line is a truncated power law fit of $\mathcal{S}_{\Phi}(\lambda)$, and the purple (dashed) line is the corner wavelength $\chi$, as determined by the fitting method described in Appendix B.}
\label{fig:froth}
\end{figure*}

For sufficiently low $R$ and over a range of $E$, we observe a spatiotemporal pattern in the $\Phi$ field: low phase \textit{patches} are separated by high-phase \textit{`fences'} (see first and fourth rows of figure \ref{fig:froth}). 
The pattern constantly shifts: cascades are more likely to occur along the `fences', relaxing the oscillators and leaving a low phase patch where a fence once stood. 
Simultaneously, cascades are unable to propagate through large patches, and instead stop in their midst - leaving a `fence' where a patch was before.
For the sake of reference, we dub this spatiotemporal behaviour \textit{froth}.

As depicted in the first and fourth rows of figure \ref{fig:froth}, the size of the patches increases along with $E$.
As explained above, each patch is the imprint of a past cascade -  and therefore patch sizes are linked to cascade sizes.
Consequently, this behaviour can also be observed in the respective complementary cumulative probability distribution (CCDF) of cascade sizes, presented in the second and fourth rows of figure \ref{fig:froth}.
Specifically, as $E$ increases, patches are enlarged and the CCDF of cascade sizes extends further towards the right.
Eventually, the CCDF forms a power law with exponent one, producing what is often referred to as Zipf's law \cite{saichev2010introduction}. 
At the same point, the probability of a global sized patch becomes nonzero.
Further increasing $E$, results in a cascade size distribution typical of supercritical dynamics. 
An example of supercritical frothing is depicted in the bottom right panel of figure \ref{fig:froth}. Note the global sized patch, connected top-to-bottom, and the supercritical CCDF.

To quantify the presence of froth, and the associated scale of the patches, we consider:
\begin{equation}
s_{\Phi}( \vec \lambda ) =\langle H_{t}^2( \vec \lambda) \rangle
\end{equation}
where $H_t(\vec \lambda)$ is the spatial Fourier transform of $\Phi_t$ for the wavelengths $\vec \lambda  = [\lambda_1,\lambda_2]$, and $\langle . \rangle$ is the average of multiple realisations over time.
We exploit the radial symmetry of the model by taking the radial mean of $ s _{\Phi}(\vec \lambda)$:
\begin{equation}
\mathcal{S} _{\Phi}( \lambda ) = \frac{1} {2 \pi |\lambda|} \int_{\Omega_\lambda} s _{\Phi}( \vec \lambda) d \Omega_\lambda
\label{mathcals}
\end{equation}
where $\Omega_\lambda$ is a circular shell of radius $\lambda = || \vec{\lambda} ||$.

Rows 3 and 6 of figure \ref{fig:froth} reveal that frothing is accompanied by a power law increase of $\mathcal{S}_{\Phi}(\lambda)$. 
The increase starts from a low limit of $\lambda$, associated with the average spatial distance between oscillators neighbouring in the two dimensional Euclidean space, 
and persists up to a wavelength where $\mathcal{S} _{\Phi}(\lambda)$ forms a `knee'. 
The wavelength associated with the `knee', dubbed \textit{corner wavelength} and denoted by $\chi$, corresponds to the largest spatial scale up to which the froth exhibits a random, self-similar symmetry, as seen in rows 1 and 4 of figure \ref{fig:froth}.

These observations allow us to numerically quantify the presence of froth and the size of the patches, by fitting a truncated power law over $\mathcal{S}_{\Phi}(\lambda)$ (see the red line on rows 3 and 6 of figure \ref{fig:froth}).
The fitting method places the power law truncation point at a wavelength that serves as an approximation for $\chi$.
Additionally, the goodness of the fit (given by $r^2 \in [0,1]$) quantifies how well $\mathcal{S}_{\Phi}(\lambda)$ follows a truncated power law.
Details on this method can be found in Appendix B. 

The metric $h$ (defined in Appendix A, equation \eqref{eq:hdef}) can be used along with $r^2$, to define \textit{empirical criteria} for the identification of  synchrony and frothing. Concretely, with $m_h, m_{r^2}$ being two threshold constants, we have:
\begin{subequations}
	\label{eq:sync}
	\begin{align}
  		h(E,R,N,\Theta) 
		& \left\{
		\begin{array}{ll}
    			> m_h,& \text{asynchrony} \\
      			\leq m_h, & \text{synchrony} \\
		\end{array} 
		\right. \\
 		r^2(E,R,N,\Theta) 
		& \left\{
		\begin{array}{ll}
			> m_{r^2},  & \text{frothing up to scale}~ \chi \\
     			 \leq m_{r^2},  & \text{no frothing}  
		\end{array} 
		\right. 
	\end{align}
\end{subequations}

\begin{figure*}[!htp]
	\includegraphics[width=2\columnwidth]{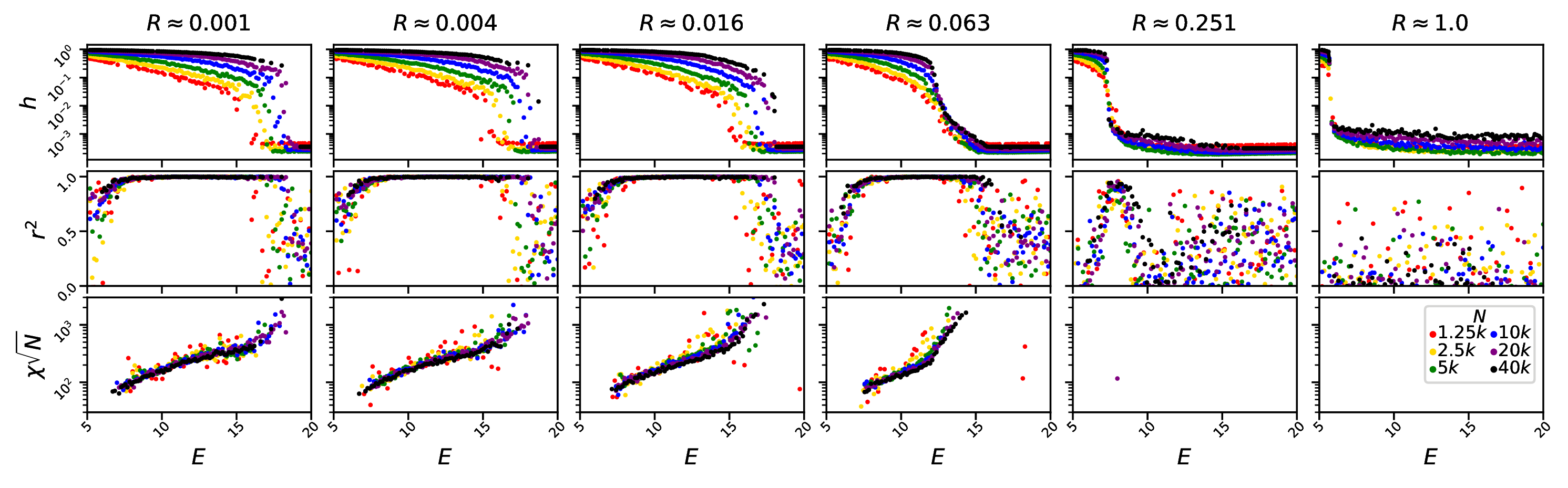}
	\captionof{figure}{Transitions in the 
	Discretised Integrate \& Fire system
	: $h,r^2$ and $\chi \sqrt{N}$ (defined in equation \eqref{eq:hdef}, Appendix B \textsection 4, and Section \RNum{3} \textsection 5)
	 as functions of the mean degree $E$, for a number of system sizes $N$, with colder colors indicating larger system sizes ($1.25k, 2.5k, 5k, 10k, 20k, 40k$).
	Long range connectivity $R$ increases according to a geometric series
, from left to right. Low values of $h$ indicate synchrony, and $r^2$ near one implies the presence of spatiotemporal patterns. Missing points in the bottom row indicate the absence of pattern formation, for the two largest $R$ values. Note the transitions i) from high to low values of $h$ and ii) ascending and descending to a plateau of $r^2$. The sharpness of both transitions increases with $N$, demonstrating that larger systems exhibit more prominent patterns and sharper transitions between macroscopic behaviours.  }
	\label{fig:FSS}
\end{figure*}

\section{Transition dynamics}
The previously proposed criteria \eqref{eq:sync} require estimates for the thresholds $m_{r^2}, m_h$ - which we obtain through the experiment presented in the current section.
$R$ and $N$ are sampled geometrically (as defined in figure \ref{fig:FSS}), while $E$ takes $75$ values evenly spaced in $[5,20]$.
The results, shown in figure \ref{fig:FSS}, reveal abrupt transitions in the macroscopic behaviour of the model, allowing us to draw three conclusions:

\paragraph{Transition to synchrony:}
In the top row of figure \ref{fig:FSS} we observe a sharp transition, from high to low $h$, revealing that the system suddenly moves from asynchrony to synchrony beyond a value of $E$.
As $R$ increases, the transition shifts to smaller $E$ values, revealing that spatial embeddedness delays the onset of synchrony.
The sharpness of the transition increases along with system size, demonstrating that this transition will still be present - and even more prominent - in the thermodynamic limit.

\paragraph{Frothing delays synchrony:}
The second row in figure \ref{fig:FSS} depicts a clear plateau of the $r^2$ metric, implying the presence of frothing dynamics over a range of $E$ values. 
The frothing regime is interposed between asynchrony and synchrony, with its width decreasing as $R$ increases.
Therefore, while high $R$ systems enter synchrony, low $R$ systems froth instead - implying that frothing is the mechanism that delays synchrony. 
The presence of the frothing regime does not depend on size, since the plateau of $r^2$ persists - and even widens - along with $N$.

\paragraph{Scaling of $\chi$:}
The bottom row of figure \ref{fig:FSS} shows that the corner frequency $\chi$ increases with $N$. 
Specifically, for the range of values in this study (N from $1.25k$ to $40k$) plotting $\chi \sqrt{N}$ makes simulation results for all $N$ to collapse into a single universal curve, revealing the presence of a scaling law.
Finally, for all panels in the bottom row of figure \ref{fig:FSS}, the peak of $\chi$ coincides with the end of the $r^2$ plateau, verifying that frothing patterns become the most prominent on the verge of synchrony.

\section{Empirical regime diagram}
\begin{figure}[htp]   
\includegraphics[width=1\columnwidth]{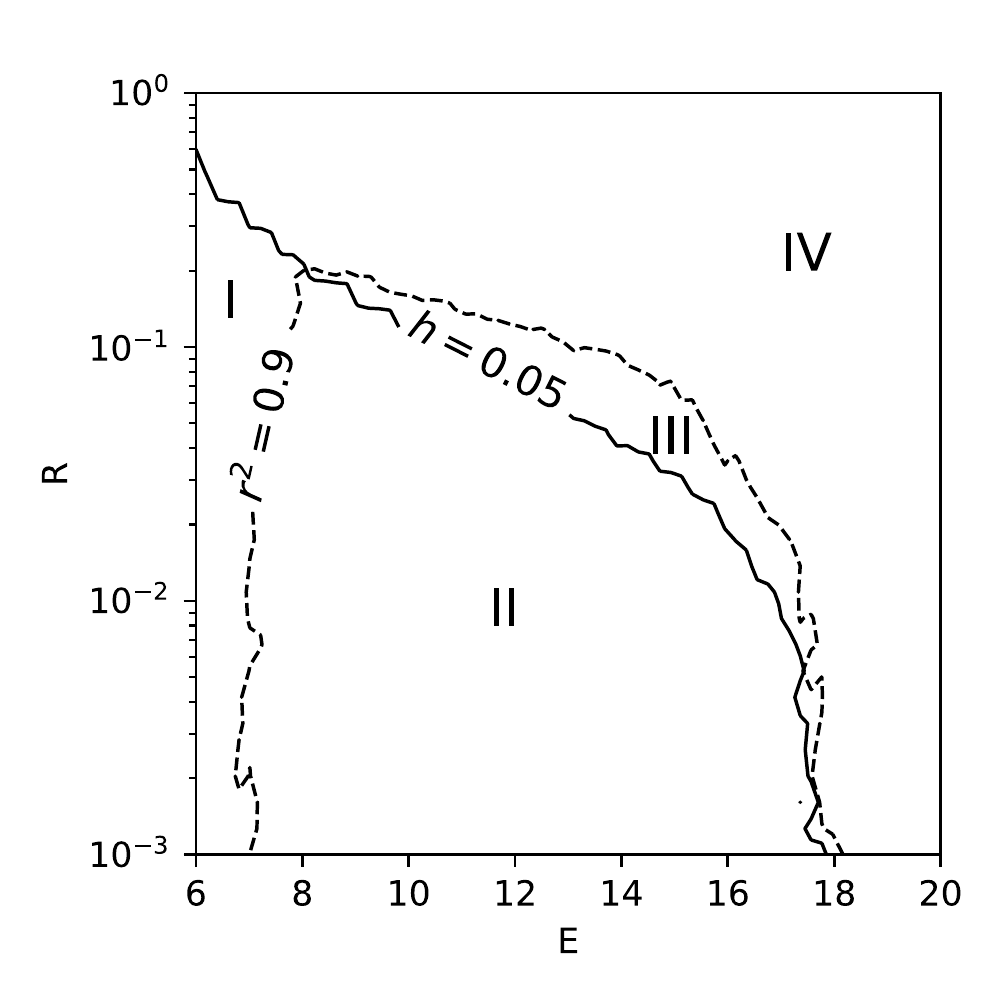}
\caption{Empirical regime diagram of the Discretised Integrate \& Fire model over the space of degree density and long range connectivity $(E,R)$. We observe four regimes: asynchrony (\RNum{1}), pattern formation (\RNum{2}), synchrony (\RNum{4}), and metastability between pattern formation and synchrony (\RNum{3}). The regime boundaries are drawn based on the criteria defined in \eqref{eq:sync}}. 
\label{fig:regimes}
\end{figure}

To numerically investigate the macroscopic behaviour of the model over the $E,R$ space, we fix $N=10^4$, and vary $R$ over $70$ evenly spaced values in $[ 6, 20 ]$. $E$ takes $30$ geometrically-spaced values in  $ [10^{-3}, 1] $. 
In order to use the criteria \eqref{eq:sync}, we need to set $m_h,m_{r^2}$. 
Visual inspection of the results in figure \ref{fig:FSS} reveals that $m_h=5~10^{-2}$ and $m_{r^2}=0.9$ allow criteria \eqref{eq:sync} to separate the macroscopic regimes adequately well for illustrative purposes.
The resulting empirical regime diagram is depicted in figure \ref{fig:regimes}, revealing four regimes:
\begin{itemize}
\item{\textit{Regime \RNum{1}:} 
For low connection density the system exhibits local cascades, with no frothing. The literature\cite{deville2008synchrony,deville2010dynamics} refers to this regime as \textit{asynchrony}.
} 
\item{\textit{Regime \RNum{2}:} 
For low long range connectivity, and over a mid-range of $E$, frothing appears in the $\Phi$ field. We refer to this regime as \textit{frothing regime}. 
}
\item{\textit{Regime \RNum{3}:}
Starting from froth and sufficiently increasing $E$ results in a regime where we intermittently observe the phenomenology of regimes \RNum{2} and \RNum{4}. The CCDF of cascade sizes is characteristic of a supercritical system, with slower than power law decay, and global-sized cascades appearing regularly. This regime is dubbed \textit{metastable regime}. 
}
\item{\textit{Regime \RNum{4}:}
For high connection density, the system undergoes a discontinuous limit cycle. 
As the average phase increases with time, cascades remain local in scale, until a global sized cascade occurs - resulting in the relaxation of the phase field.  
Then, the buildup of phase synchronisation begins anew.
We refer to this regime as \textit{synchrony}. 
} 
\end{itemize}

\section{Discussion}
We have numerically investigated the behaviour of discretised Integrate \& Fire oscillators, with slow stochastic drive, over spatial graphs. 
Remarkably, when placed over spatial topologies these models give rise to novel, nontrivial dynamics:
spatiotemporal patterns form, where large clusters of relaxed nodes act as natural barriers against cascading, while jagged strips of near-firing nodes facilitate long distance propagation of cascades.
This pattern can give rise to critical dynamics with cascading events sizes following a power law with exponent one. 
Further increasing the number of edges results in metastability between pattern formation and synchrony, and eventually drives the model into synchrony.

Our findings show that taking into consideration the fundamentally spatial character of neuronal networks drastically extends the phenomenological overlap between discretised Integrate \& Fire and more intricate neuronal models. 
This observation provides solid grounds for the usage of the discretised Integrate \& Fire model in the study of pattern formation and critical dynamics.
Further numerical studies could shed light on the role of additional neuronal properties (leakiness, nonlinear response curves, and inhibition) on pattern formation and criticality in neuronal systems.

In spite of the extensive simulations in the current study, the exact mathematical nature of the regime transitions of the model is not well understood - and could constitute the subject of future works. 
Additionally, to the best of the authors' knowledge, frothing has not been so far empirically observed in physical systems. 
This is surprising considering the generality of the conditions under which frothing arises.
A possible explanation could be that detecting froth requires knowledge of the \textit{phase}, an attribute of oscillators that is typically less prominent and harder to measure than their  \textit{state}.
In any case, further works are warranted, focusing on the empirical detection of frothing in Pulse Coupled Oscillator systems.

One of the authors (DG) would like to thank S.C. Lera and S.O.P. Blume for stimulating discussions, and is supported by the Future Resilient Systems at the Singapore-ETH Centre (SEC), a collaboration between ETH Zurich and National Research Foundation (NRF) of Singapore (FI 370074011) - under the auspices of the NRF's Campus for Research Excellence and Technological Enterprise (CREATE) programme. 

\appendix
\section{Quantifying synchrony}

The near-periodic behaviour of the time series of cascade sizes $C$ can be detected in the frequency domain, where peaks appear in the power spectrum.
The concentration of the spectral power around these peaks is quantified using the Herfindahl-Hirschman index. 
Since the data are discrete, we will be using the discrete Fourier transform.
Let $\mathcal{P}_{C}(  f )$ be the temporal \textit{spectral power} of $C$ for frequency $f$:
\setlength{\jot}{0pt}
\begin{gather} \hat h = \sum_f 	\left(
	 \frac 
	{    \mathcal{P}_{C}(f)     } 
	{      \mathcal{N}      }
	\right) ^2 ,
	~~~~ \mathcal{N} = \sum_f  \mathcal{P}_{C}(f) \\
	 h = \frac {    \hat h    -  n_s     } {      1- n_s }
	 \label{eq:hdef}
\end{gather}
The $\hat h$ metric takes values in $[0,1]$ and quantifies the concentration of $\mathcal{P}_{C}(f)$ around a few frequencies. 
During asynchrony, the only significantly contributing frequency is the $0$th, resulting in near zero values of $\mathcal{P}_{C}(f)$ for $f>0$ and therefore $h$ near one. 
In contrast, during synchrony, higher harmonics carry considerable amount of power resulting in a lower $h$ index. 
Discerning between synchrony and asynchrony does require a threshold value of $h$, which we determine empirically in this study to be equal to $5 \cdot 10^{-2}$ (see Section \RNum{5} \textsection1 for details).

\section{Quantifying spatiotemporal pattern formation}

We observe that in the case of frothing patterns, $\mathcal{ S }_{\Phi} (\lambda)$ follows a power law increase, over a band of wavelengths.
The upper limit of the band $\chi$ is associated with the largest cell size  of the frothing pattern. 
The lower limit of the band is proportional to the size of the mesh used to estimate the $\Phi$ field.
In the current study, $\sqrt{N}$ oscillators are placed along each dimension of the two-dimensional Euclidean space, resulting in an lower wavelength limit of $4 \pi / \sqrt{N} $.

Higher frequency components also reside in the $\Phi$ field, for example due to the randomness in the placement of the oscillators.
These higher frequency components may introduce noise in the spatial spectrum of $\Phi$, through a processes known as \textit{aliasing}.
Specifically in the case of data produced by processes with power law spectra, aliasing results in the measured spectrum progressively resembling white noise as we move to smaller wavelengths\cite{kirchner2005aliasing}. 
As a treatment, we ignore the values of the measured spectrum for small wavelengths, by doubling the lower limit derived in the previous paragraph to $\lambda_{\text{min}}=8 \pi / \sqrt{N}$. 

To determine the corner wavelength $\chi$, we consider the frequency response function of a linear low pass filter:
\begin{equation}
g(\lambda;p_1, p_2, p_3, p_4) = \frac{p_1} {\sqrt{1 + \big( \lambda / p_3 \big)^{-2p_4} } } +p_2
\label{eq:fit}
\end{equation}
and fit it to $\mathcal{ S }_{\Phi}(\lambda)$, according to the following equation:
\setlength{\jot}{0pt}
\begin{multline}
(p_1^*, p_2^*, p_3^*, p_4^*) =\\ 
\argmin_{ \substack{p_1,p_2,\\p_3,p_4}}  
~ \sum_{ \lambda \geq \lambda_{\text{min}}} \left( \frac{ \mathcal{ S }_{\Phi}(\lambda) -  g(\lambda;p_1,p_2,p_3,p_4)} 
{ \mathcal{ S }_{\Phi}(\lambda)}
 \right)^2
\label{eq:fit_opt}
\end{multline}
where $p_1, p_2, p_3,p_4$ are fitting parameters.
Equation \eqref{eq:fit} describes a power law increase with exponent $p_4$, up to the wavelength $p_3$ where the function forms a visual `knee'.  
From that point onwards, the value of \eqref{eq:fit} remains nearly constant at $p_1+p_2$.
For the sake of illustration, and to enable comparison between different  $\mathcal{ S }_{\Phi}(\lambda)$ curves, we position the truncation point $\chi$ at $ p_3^*$.

Initial solutions to the problem in \eqref{eq:fit_opt} were obtained via the particle swarm method, and refined with a Levenberg-Marquardt local search.
Examples of the fitted $g(\lambda;p_1^*, p_2^*, p_3^*, p_4^*) $ are depicted in rows 3 and 6 of Figure \ref{fig:froth} (red solid line), along with the corresponding $\mathcal{ S }_{\Phi}(\lambda)$ values.
The quality of the fits can be assessed via the r-squared metric, to which we will refer to as $r^2$. 
Specifically, $r^2$ is used to determine whether a simulation exhibits frothing: 
high value of $r^2$ implies that equation \ref{eq:fit} provides a good fit, providing evidence in support of the presence of froth in the $\Phi$ field.


\bibliography{biblio}

\begin{thebibliography}{21}%
\makeatletter
\providecommand \@ifxundefined [1]{%
 \@ifx{#1\undefined}
}%
\providecommand \@ifnum [1]{%
 \ifnum #1\expandafter \@firstoftwo
 \else \expandafter \@secondoftwo
 \fi
}%
\providecommand \@ifx [1]{%
 \ifx #1\expandafter \@firstoftwo
 \else \expandafter \@secondoftwo
 \fi
}%
\providecommand \natexlab [1]{#1}%
\providecommand \enquote  [1]{``#1''}%
\providecommand \bibnamefont  [1]{#1}%
\providecommand \bibfnamefont [1]{#1}%
\providecommand \citenamefont [1]{#1}%
\providecommand \href@noop [0]{\@secondoftwo}%
\providecommand \href [0]{\begingroup \@sanitize@url \@href}%
\providecommand \@href[1]{\@@startlink{#1}\@@href}%
\providecommand \@@href[1]{\endgroup#1\@@endlink}%
\providecommand \@sanitize@url [0]{\catcode `\\12\catcode `\$12\catcode
  `\&12\catcode `\#12\catcode `\^12\catcode `\_12\catcode `\%12\relax}%
\providecommand \@@startlink[1]{}%
\providecommand \@@endlink[0]{}%
\providecommand \url  [0]{\begingroup\@sanitize@url \@url }%
\providecommand \@url [1]{\endgroup\@href {#1}{\urlprefix }}%
\providecommand \urlprefix  [0]{URL }%
\providecommand \Eprint [0]{\href }%
\providecommand \doibase [0]{http://dx.doi.org/}%
\providecommand \selectlanguage [0]{\@gobble}%
\providecommand \bibinfo  [0]{\@secondoftwo}%
\providecommand \bibfield  [0]{\@secondoftwo}%
\providecommand \translation [1]{[#1]}%
\providecommand \BibitemOpen [0]{}%
\providecommand \bibitemStop [0]{}%
\providecommand \bibitemNoStop [0]{.\EOS\space}%
\providecommand \EOS [0]{\spacefactor3000\relax}%
\providecommand \BibitemShut  [1]{\csname bibitem#1\endcsname}%
\let\auto@bib@innerbib\@empty
\bibitem [{\citenamefont {Mirollo}\ and\ \citenamefont
  {Strogatz}(1990)}]{mirollo1990synchronization}%
  \BibitemOpen
  \bibfield  {author} {\bibinfo {author} {\bibfnamefont {R.~E.}\ \bibnamefont
  {Mirollo}}\ and\ \bibinfo {author} {\bibfnamefont {S.~H.}\ \bibnamefont
  {Strogatz}},\ }\href@noop {} {\bibfield  {journal} {\bibinfo  {journal} {SIAM
  Journal on Applied Mathematics}\ }\textbf {\bibinfo {volume} {50}},\ \bibinfo
  {pages} {1645} (\bibinfo {year} {1990})}\BibitemShut {NoStop}%
\bibitem [{\citenamefont {Proskurnikov}\ and\ \citenamefont
  {Cao}(2017)}]{proskurnikov2017synchronization}%
  \BibitemOpen
  \bibfield  {author} {\bibinfo {author} {\bibfnamefont {A.~V.}\ \bibnamefont
  {Proskurnikov}}\ and\ \bibinfo {author} {\bibfnamefont {M.}~\bibnamefont
  {Cao}},\ }\href@noop {} {\bibfield  {journal} {\bibinfo  {journal} {IEEE
  Transactions on Automatic Control}\ }\textbf {\bibinfo {volume} {62}},\
  \bibinfo {pages} {5873} (\bibinfo {year} {2017})}\BibitemShut {NoStop}%
\bibitem [{\citenamefont {Konishi}\ and\ \citenamefont
  {Kokame}(2008)}]{konishi2008synchronization}%
  \BibitemOpen
  \bibfield  {author} {\bibinfo {author} {\bibfnamefont {K.}~\bibnamefont
  {Konishi}}\ and\ \bibinfo {author} {\bibfnamefont {H.}~\bibnamefont
  {Kokame}},\ }\href@noop {} {\bibfield  {journal} {\bibinfo  {journal} {Chaos:
  An Interdisciplinary Journal of Nonlinear Science}\ }\textbf {\bibinfo
  {volume} {18}},\ \bibinfo {pages} {033132} (\bibinfo {year}
  {2008})}\BibitemShut {NoStop}%
\bibitem [{\citenamefont {Wray}\ and\ \citenamefont
  {Bishop}(2014)}]{wray2014cascades}%
  \BibitemOpen
  \bibfield  {author} {\bibinfo {author} {\bibfnamefont {C.~M.}\ \bibnamefont
  {Wray}}\ and\ \bibinfo {author} {\bibfnamefont {S.~R.}\ \bibnamefont
  {Bishop}},\ }\href@noop {} {\bibfield  {journal} {\bibinfo  {journal}
  {Scientific Reports}\ }\textbf {\bibinfo {volume} {4}},\ \bibinfo {pages}
  {6355} (\bibinfo {year} {2014})}\BibitemShut {NoStop}%
\bibitem [{\citenamefont {Wray}\ and\ \citenamefont
  {Bishop}(2016)}]{wray2016financial}%
  \BibitemOpen
  \bibfield  {author} {\bibinfo {author} {\bibfnamefont {C.~M.}\ \bibnamefont
  {Wray}}\ and\ \bibinfo {author} {\bibfnamefont {S.~R.}\ \bibnamefont
  {Bishop}},\ }\href@noop {} {\bibfield  {journal} {\bibinfo  {journal} {PloS
  One}\ }\textbf {\bibinfo {volume} {11}},\ \bibinfo {pages} {e0151790}
  (\bibinfo {year} {2016})}\BibitemShut {NoStop}%
\bibitem [{\citenamefont {DeVille}\ and\ \citenamefont
  {Peskin}(2008)}]{deville2008synchrony}%
  \BibitemOpen
  \bibfield  {author} {\bibinfo {author} {\bibfnamefont {R.~L.}\ \bibnamefont
  {DeVille}}\ and\ \bibinfo {author} {\bibfnamefont {C.~S.}\ \bibnamefont
  {Peskin}},\ }\href@noop {} {\bibfield  {journal} {\bibinfo  {journal}
  {Bulletin of Mathematical Biology}\ }\textbf {\bibinfo {volume} {70}},\
  \bibinfo {pages} {1608} (\bibinfo {year} {2008})}\BibitemShut {NoStop}%
\bibitem [{\citenamefont {Huang}\ \emph {et~al.}(2004)\citenamefont {Huang},
  \citenamefont {Troy}, \citenamefont {Yang}, \citenamefont {Ma}, \citenamefont
  {Laing}, \citenamefont {Schiff},\ and\ \citenamefont {Wu}}]{huang2004spiral}%
  \BibitemOpen
  \bibfield  {author} {\bibinfo {author} {\bibfnamefont {X.}~\bibnamefont
  {Huang}}, \bibinfo {author} {\bibfnamefont {W.~C.}\ \bibnamefont {Troy}},
  \bibinfo {author} {\bibfnamefont {Q.}~\bibnamefont {Yang}}, \bibinfo {author}
  {\bibfnamefont {H.}~\bibnamefont {Ma}}, \bibinfo {author} {\bibfnamefont
  {C.~R.}\ \bibnamefont {Laing}}, \bibinfo {author} {\bibfnamefont {S.~J.}\
  \bibnamefont {Schiff}}, \ and\ \bibinfo {author} {\bibfnamefont {J.-Y.}\
  \bibnamefont {Wu}},\ }\href@noop {} {\bibfield  {journal} {\bibinfo
  {journal} {Journal of Neuroscience}\ }\textbf {\bibinfo {volume} {24}},\
  \bibinfo {pages} {9897} (\bibinfo {year} {2004})}\BibitemShut {NoStop}%
\bibitem [{\citenamefont {Wilson}\ and\ \citenamefont
  {Cowan}(1972)}]{wilson1972excitatory}%
  \BibitemOpen
  \bibfield  {author} {\bibinfo {author} {\bibfnamefont {H.~R.}\ \bibnamefont
  {Wilson}}\ and\ \bibinfo {author} {\bibfnamefont {J.~D.}\ \bibnamefont
  {Cowan}},\ }\href@noop {} {\bibfield  {journal} {\bibinfo  {journal}
  {Biophysical Journal}\ }\textbf {\bibinfo {volume} {12}},\ \bibinfo {pages}
  {1} (\bibinfo {year} {1972})}\BibitemShut {NoStop}%
\bibitem [{\citenamefont {DeVille}\ \emph {et~al.}(2005)\citenamefont
  {DeVille}, \citenamefont {Vanden-Eijnden},\ and\ \citenamefont
  {Muratov}}]{deville2005two}%
  \BibitemOpen
  \bibfield  {author} {\bibinfo {author} {\bibfnamefont {R.~L.}\ \bibnamefont
  {DeVille}}, \bibinfo {author} {\bibfnamefont {E.}~\bibnamefont
  {Vanden-Eijnden}}, \ and\ \bibinfo {author} {\bibfnamefont {C.~B.}\
  \bibnamefont {Muratov}},\ }\href@noop {} {\bibfield  {journal} {\bibinfo
  {journal} {Physical Review E}\ }\textbf {\bibinfo {volume} {72}},\ \bibinfo
  {pages} {031105} (\bibinfo {year} {2005})}\BibitemShut {NoStop}%
\bibitem [{\citenamefont {Ermentrout}\ and\ \citenamefont
  {Kleinfeld}(2001)}]{ermentrout2001traveling}%
  \BibitemOpen
  \bibfield  {author} {\bibinfo {author} {\bibfnamefont {G.~B.}\ \bibnamefont
  {Ermentrout}}\ and\ \bibinfo {author} {\bibfnamefont {D.}~\bibnamefont
  {Kleinfeld}},\ }\href@noop {} {\bibfield  {journal} {\bibinfo  {journal}
  {Neuron}\ }\textbf {\bibinfo {volume} {29}},\ \bibinfo {pages} {33} (\bibinfo
  {year} {2001})}\BibitemShut {NoStop}%
\bibitem [{\citenamefont {Guardiola}\ and\ \citenamefont
  {D\'{\i}az-Guilera}(1999)}]{Guardiola99}%
  \BibitemOpen
  \bibfield  {author} {\bibinfo {author} {\bibfnamefont {X.}~\bibnamefont
  {Guardiola}}\ and\ \bibinfo {author} {\bibfnamefont {A.}~\bibnamefont
  {D\'{\i}az-Guilera}},\ }\href {\doibase 10.1103/PhysRevE.60.3626} {\bibfield
  {journal} {\bibinfo  {journal} {Phys. Rev. E}\ }\textbf {\bibinfo {volume}
  {60}},\ \bibinfo {pages} {3626} (\bibinfo {year} {1999})}\BibitemShut
  {NoStop}%
\bibitem [{\citenamefont {Bottani}(1995)}]{bottani1995pulse}%
  \BibitemOpen
  \bibfield  {author} {\bibinfo {author} {\bibfnamefont {S.}~\bibnamefont
  {Bottani}},\ }\href@noop {} {\bibfield  {journal} {\bibinfo  {journal}
  {Physical Review Letters}\ }\textbf {\bibinfo {volume} {74}},\ \bibinfo
  {pages} {4189} (\bibinfo {year} {1995})}\BibitemShut {NoStop}%
\bibitem [{\citenamefont {Friedman}\ \emph {et~al.}(2012)\citenamefont
  {Friedman}, \citenamefont {Ito}, \citenamefont {Brinkman}, \citenamefont
  {Shimono}, \citenamefont {DeVille}, \citenamefont {Dahmen}, \citenamefont
  {Beggs},\ and\ \citenamefont {Butler}}]{friedman2012universal}%
  \BibitemOpen
  \bibfield  {author} {\bibinfo {author} {\bibfnamefont {N.}~\bibnamefont
  {Friedman}}, \bibinfo {author} {\bibfnamefont {S.}~\bibnamefont {Ito}},
  \bibinfo {author} {\bibfnamefont {B.~A.}\ \bibnamefont {Brinkman}}, \bibinfo
  {author} {\bibfnamefont {M.}~\bibnamefont {Shimono}}, \bibinfo {author}
  {\bibfnamefont {R.~L.}\ \bibnamefont {DeVille}}, \bibinfo {author}
  {\bibfnamefont {K.~A.}\ \bibnamefont {Dahmen}}, \bibinfo {author}
  {\bibfnamefont {J.~M.}\ \bibnamefont {Beggs}}, \ and\ \bibinfo {author}
  {\bibfnamefont {T.~C.}\ \bibnamefont {Butler}},\ }\href@noop {} {\bibfield
  {journal} {\bibinfo  {journal} {Physical Review Letters}\ }\textbf {\bibinfo
  {volume} {108}},\ \bibinfo {pages} {208102} (\bibinfo {year}
  {2012})}\BibitemShut {NoStop}%
\bibitem [{\citenamefont {DeVille}\ and\ \citenamefont
  {Peskin}(2012)}]{deville2012synchrony}%
  \BibitemOpen
  \bibfield  {author} {\bibinfo {author} {\bibfnamefont {R.~L.}\ \bibnamefont
  {DeVille}}\ and\ \bibinfo {author} {\bibfnamefont {C.~S.}\ \bibnamefont
  {Peskin}},\ }\href@noop {} {\bibfield  {journal} {\bibinfo  {journal}
  {Bulletin of Mathematical Biology}\ }\textbf {\bibinfo {volume} {74}},\
  \bibinfo {pages} {769} (\bibinfo {year} {2012})}\BibitemShut {NoStop}%
\bibitem [{\citenamefont {Beggs}\ and\ \citenamefont
  {Plenz}(2003)}]{beggs2003neuronal}%
  \BibitemOpen
  \bibfield  {author} {\bibinfo {author} {\bibfnamefont {J.~M.}\ \bibnamefont
  {Beggs}}\ and\ \bibinfo {author} {\bibfnamefont {D.}~\bibnamefont {Plenz}},\
  }\href@noop {} {\bibfield  {journal} {\bibinfo  {journal} {Journal of
  Neuroscience}\ }\textbf {\bibinfo {volume} {23}},\ \bibinfo {pages} {11167}
  (\bibinfo {year} {2003})}\BibitemShut {NoStop}%
\bibitem [{\citenamefont {Kinouchi}\ and\ \citenamefont
  {Copelli}(2006)}]{kinouchi2006optimal}%
  \BibitemOpen
  \bibfield  {author} {\bibinfo {author} {\bibfnamefont {O.}~\bibnamefont
  {Kinouchi}}\ and\ \bibinfo {author} {\bibfnamefont {M.}~\bibnamefont
  {Copelli}},\ }\href@noop {} {\bibfield  {journal} {\bibinfo  {journal}
  {Nature Physics}\ }\textbf {\bibinfo {volume} {2}},\ \bibinfo {pages} {348}
  (\bibinfo {year} {2006})}\BibitemShut {NoStop}%
\bibitem [{\citenamefont {Beggs}(2008)}]{beggs2008criticality}%
  \BibitemOpen
  \bibfield  {author} {\bibinfo {author} {\bibfnamefont {J.~M.}\ \bibnamefont
  {Beggs}},\ }\href@noop {} {\bibfield  {journal} {\bibinfo  {journal}
  {Philosophical Transactions of the Royal Society of London A: Mathematical,
  Physical and Engineering Sciences}\ }\textbf {\bibinfo {volume} {366}},\
  \bibinfo {pages} {329} (\bibinfo {year} {2008})}\BibitemShut {NoStop}%
\bibitem [{\citenamefont {Penrose}\ \emph {et~al.}(2003)\citenamefont {Penrose}
  \emph {et~al.}}]{penrose2003random}%
  \BibitemOpen
  \bibfield  {author} {\bibinfo {author} {\bibfnamefont {M.}~\bibnamefont
  {Penrose}} \emph {et~al.},\ }\href@noop {} {\emph {\bibinfo {title} {Random
  geometric graphs}}},\ \bibinfo {number} {5}\ (\bibinfo  {publisher} {Oxford
  university press},\ \bibinfo {year} {2003})\BibitemShut {NoStop}%
\bibitem [{\citenamefont {Saichev}\ \emph {et~al.}(2010)\citenamefont
  {Saichev}, \citenamefont {Malevergne},\ and\ \citenamefont
  {Sornette}}]{saichev2010introduction}%
  \BibitemOpen
  \bibfield  {author} {\bibinfo {author} {\bibfnamefont {A.}~\bibnamefont
  {Saichev}}, \bibinfo {author} {\bibfnamefont {Y.}~\bibnamefont {Malevergne}},
  \ and\ \bibinfo {author} {\bibfnamefont {D.}~\bibnamefont {Sornette}},\
  }\href@noop {} {\emph {\bibinfo {title} {Theory of Zipf's Law and Beyond}}}\
  (\bibinfo  {publisher} {Springer},\ \bibinfo {year} {2010})\BibitemShut
  {NoStop}%
\bibitem [{\citenamefont {DeVille}\ \emph {et~al.}(2010)\citenamefont
  {DeVille}, \citenamefont {Peskin},\ and\ \citenamefont
  {Spencer}}]{deville2010dynamics}%
  \BibitemOpen
  \bibfield  {author} {\bibinfo {author} {\bibfnamefont {R.~L.}\ \bibnamefont
  {DeVille}}, \bibinfo {author} {\bibfnamefont {C.~S.}\ \bibnamefont {Peskin}},
  \ and\ \bibinfo {author} {\bibfnamefont {J.~H.}\ \bibnamefont {Spencer}},\
  }\href@noop {} {\bibfield  {journal} {\bibinfo  {journal} {Mathematical
  Modelling of Natural Phenomena}\ }\textbf {\bibinfo {volume} {5}},\ \bibinfo
  {pages} {26} (\bibinfo {year} {2010})}\BibitemShut {NoStop}%
\bibitem [{\citenamefont {Kirchner}(2005)}]{kirchner2005aliasing}%
  \BibitemOpen
  \bibfield  {author} {\bibinfo {author} {\bibfnamefont {J.~W.}\ \bibnamefont
  {Kirchner}},\ }\href@noop {} {\bibfield  {journal} {\bibinfo  {journal}
  {Physical Review E}\ }\textbf {\bibinfo {volume} {71}} (\bibinfo {year}
  {2005})}\BibitemShut {NoStop}%
\end{thebibliography}%

\end{document}